Title: Cluster-based name embeddings reduce ethnic disparities in record linkage quality under realistic name corruption: evidence from the North Carolina Voter Registry

Running head: Name embeddings and ethnic linkage disparities

Authors:

Joseph Lam[1], Mario Cortina-Borja[1], Rob Aldridge[2], Ruth Blackburn[1], Katie Harron[1]

1 Great Ormond Street Institute of Child Health, University College London, London, UK

2 Institute for Health Metrics and Evaluation, University of Washington, Seattle, USA

Word count: 2985

**Abstract**

**Background:** Differential ethnic-based record linkage errors can bias epidemiologic estimates. Prior evidence often conflates heterogeneity in error mechanisms with unequal exposure to error.

**Methods:** Using snapshots of the North Carolina Voter Registry (Oct 2011–Oct 2022), we derived empirical name-discrepancy profiles to parameterise realistic corruptions. From an Oct 2022 extract (n=848,566), we generated five replicate corrupted datasets under three settings that separately varied mechanism heterogeneity and exposure inequality, and linked records back to originals using unadjusted Jaro–Winkler, Term Frequency (TF)-adjusted Jaro–Winkler, and a cluster-based forename-embedding comparator combined with TF-adjusted surname comparison. We evaluated false match rate (FMR), missed match rate (MMR) and white-centric disparities.

**Results:** At a fixed MMR near 0.20, overall error rates and ethnic disparities diverged substantially by model under disproportionate exposure to corruption. Term-frequency (TF)-adjusted Jaro–Winkler achieved very low overall FMR (0.55% (95% CI 0.54–0.57)) at overall MMR 20.34% (20.30–20.39), but large white-centric under-linkage disparities persisted: Hispanic voters had 36.3% (36.1–36.6) and Non-Hispanic Black voters 8.6% (8.6–8.7) higher FMRs compared to Non-Hispanic White groups. Relative to unadjusted string similarity, TF adjustment reduced these disparities (Hispanic: +60.4% (60.1–60.7) to +36.3%; Black: +13.1% (13.0–13.2) to +8.6%). The cluster-based forename-embedding model reduced missed-match disparities further (Hispanic: +10.2% (9.8–10.3); Black: +0.6% (0.4–0.7)), but at a cost of increasing overall FMR (4.28% (4.22–4.35)) at the same threshold.

**Conclusions:** Unequal exposure to identifier error drove substantially larger disparities than mechanism heterogeneity alone; cluster-based embeddings markedly narrowed under-linkage disparities beyond TF adjustment.

## Key Messages

- We sought to demonstrate how ethnic disparities in linkage error reflect both how identifiers vary (mechanisms) and who is more often affected (exposure), and their interaction with linkage model design can amplify or attenuate inequities.

- Term-frequency adjustment provides partial mitigation of under-linkage disparities relative to unadjusted string similarity, while our novel cluster-based name embeddings model substantially narrows these disparities, but at a cost of higher false matches.

- Linkage evaluations should report subgroup-specific error exposure and mechanisms, alongside linkage quality metrics, which allow a more transparent evaluation of how potential differential linkage bias are introduced.

## Introduction

Record linkage has become a core enabling mechanism for epidemiologic research using administrative data, supporting longitudinal follow-up, cohort construction, and enriched covariate ascertainment without new primary data collection (1). Yet linkage quality is rarely uniform across populations. When linkage error varies systematically by groups, it can induce differential misclassification and selection, biasing estimates of disease burden and health inequalities—particularly in settings where linkage relies on identifiers that are socially patterned (2). A growing body of work reports variation in linkage performance by ethnicity, race, and migration status, but empirical evidence on *why* these disparities arise remains limited, and methodological guidance on targeted remedies is correspondingly underdeveloped (3–5).

Two mechanisms are often conflated in discussions of ethnic differences in linkage quality: heterogeneity in error mechanisms (differences in the form of identifier variation, e.g., truncation, spacing conventions, token order, or romanization (6,7)), and unequal exposure to error (differences in the rate of missingness, recording problems or other data defects. Distinguishing these mechanisms is not just a semantic exercise as they imply different interventions. Unequal exposure to errors points toward improvements in data capture, governance, and institutional practice; heterogeneity in error mechanisms points toward modifications to linkage models, feature engineering, and calibration strategies that are robust to structurally different identifier distributions. Disparities are frequently attributed to a generic notion of "poorer data quality" among ethnically minoritised groups (8,9), but this framing is incomplete unless it specifies *which* mechanism is operating and *how* it translates into linkage error. In practice, mechanism heterogeneity and exposure inequality only become differential false matches and missed matches through the linkage system's evidential assumptions—blocking, similarity functions and evidence weighting—which determine what kinds of variation are treated as plausible within-person change versus treated as evidence of distinct individuals. As a result, two populations can experience identical error rates but different error *types* (mechanism heterogeneity), and a given linkage approach may be robust to one pattern of variation but fragile to another. This is particularly salient for names, where institutional conventions such as truncation, token order, spacing and romanisation create structured forms of within-person variation, and where between-person ambiguity is governed by highly skewed, socially patterned name frequency distributions (7,10,11). A mechanisms-based evaluation of equity in linkage therefore requires modelling both the generation of identifier errors (mechanism and exposure) and the way linkage models convert those errors into subgroup-specific false match and missed match rates.

Within the Fellegi–Sunter framework, name agreement can be over-weighted for highly prevalent names that provide limited discriminative information and under-weighted for rarer, more identifying names (12). A common remedy is term-frequency (TF) adjustment, which downweights agreement on common name terms and upweights agreement on rare terms using inverse prevalence weights (13,14). Although TF adjustment is widely recommended and often improves aggregate discrimination, there is limited empirical evidence that it reduces ethnic disparities in linkage error, rather than improving overall precision (3).

In this study, we use repeated annual observations from the North Carolina Voter Registry (15) (NCVR) to (i) characterise heterogeneity in name structure by ethnic group, (ii) parameterise empirically grounded name corruption processes that separate heterogeneity in error mechanisms from inequality in exposure to error, and (iii) evaluate whether a novel cluster-based name-embedding linkage model reduces ethnic disparities relative to an unadjusted string-similarity baseline and a TF-adjusted baseline. We evaluate these approaches in a linkage setting that uses names alongside routinely available demographic identifiers, and we quantify both overall error rates and white-centric disparities in false matches and missed matches. By perturbing names while keeping non-name identifiers unchanged across scenarios, we attribute changes in disparity primarily to differences in (i) error mechanism structure and (ii) exposure, and to how alternative linkage models translate these into subgroup-specific error.

# Methods

## Data source and study period

We used annual October snapshots of the publicly available North Carolina Voter Registry, covering October 2011 to October 2022. The registry includes a stable voter identifier, self-reported year of birth, gender, forename and surname, race and ethnicity (combined as an 8-category ethnic group variable, which we used as the primary stratification). Ethnic categories included Asian, Hispanic (White or Black), Indigenous or Pacific Islander, Mixed, Non-Hispanic Black, Non-Hispanic White, Other and Unknown. We treated the stable voter identifier as the gold-standard linkage key for within-person comparisons (Stages 1–2) and for evaluating probabilistic linkage outputs in the linkage experiments (Stage 4).

## Stage 1: empirical within-person name discrepancies and error profiling

For each of the 10 adjacent annual snapshot pairs (2011–2012 through 2021–2022), we deterministically linked records using the stable voter identifier and identified within-person discrepancies in forename and surname between years (Figure 1). These discrepancies provide high-precision empirical examples of the mixture of true name changes and administrative recording variation relevant to linkage.

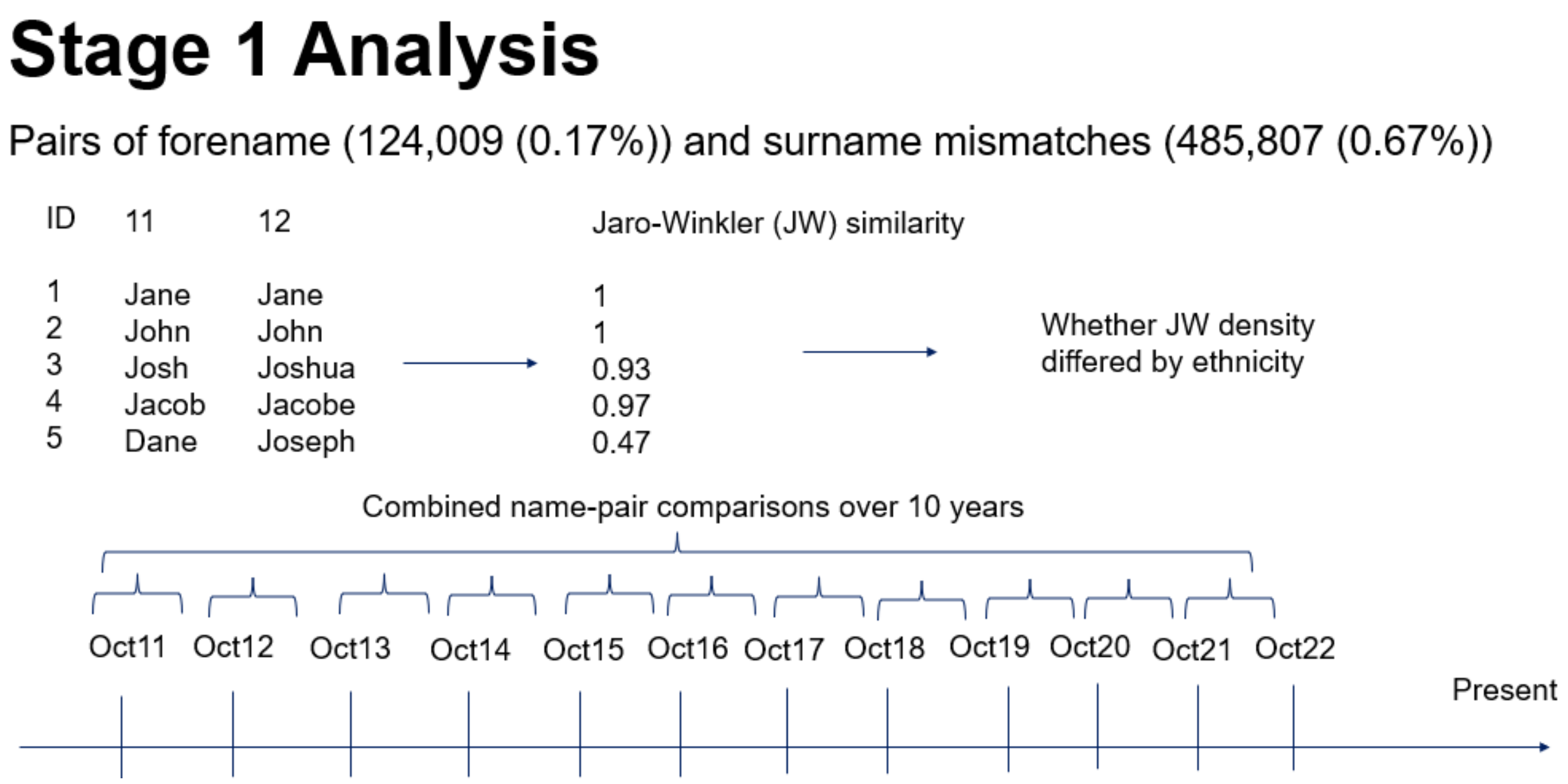

Figure 1. Study design and Stage 1 empirical discrepancy profiling across annual North Carolina Voter Register snapshots (October 2011–October 2022).

For each discrepant name pair, we computed Jaro–Winkler similarity and Levenshtein edit distance (16–18). We decomposed discrepancies into edit-operation patterns (insertion, deletion and substitution, including combinations) and recorded the approximate position of edits (start, first half, second half, end, or distributed). Stratifying these summaries by ethnicity yielded empirical error profiles used to parameterise the synthetic corruption settings in Stage 3 while preserving dependencies between edit distance, edit type and position (Supplementary Table 2).

## Stage 2: name characteristics and embedding representation

To quantify heterogeneity in name structure, we derived name-characteristic features intended to capture properties that influence ambiguity and comparator behaviour. Features included: character length (with and without spaces), number of terms, mean characters per term, vowel–consonant ratio, whether ending with a vowel, hyphen, apostrophe, and middle-name presence, counts of other names sharing the same initial trigram and distinctiveness (e.g., prevalence-based uniqueness).

We standardised features and used principal components analysis (PCA) to obtain a low-dimensional embedding of forename structure. These embeddings were used to operationalise a cluster-based forename comparison. Candidate pairs that are close in the embedding space are treated as more similar, even when superficial string edits differ. The full feature set and PCA specification are reported in Supplementary Tables 3, 4.

## Stage 3: synthetic corruption settings to separate error mechanisms from exposure

We generated five replicate corrupted versions of the October 2022 extract (n=848,566) under three corruption settings designed to separate heterogeneity in error mechanisms from inequality in exposure to error. Corruptions were applied independently to forename and surname fields. Each corruption event was generated in two steps: (1) selecting which records to corrupt (exposure) and (2) sampling the corruption form (mechanism) as a joint profile of edit distance, edit type and edit position. Mechanism profiles were sampled from the empirical Stage 1 mismatch distributions (pooled or ethnic-specific, depending on setting), preserving observed dependencies between distance, type and position.

**Setting 1 (uniform random errors).** Records were selected for corruption at a constant probability across the full sample, targeting 10% corruption of forenames and 10% corruption of surnames overall. Corruption profiles (distance/type/position) were sampled from the pooled empirical distribution across ethnic groups.

**Setting 2 (ethnic-specific mechanisms with equal exposure).** We enforced the same overall corruption rate within each ethnic group (equal exposure), while sampling corruption profiles from ethnic-specific empirical distributions. This isolates mechanism heterogeneity while holding exposure constant across groups.

**Setting 3 (disproportionate exposure).** We allocated a fixed total number of corruptions (10% overall) disproportionately across ethnic groups (unequal exposure), while retaining ethnic - specific corruption profiles. Exposure allocation used pre-specified relative weights by ethnic groups (Supplementary Table 1), producing higher corruption probability for minoritised groups and lower probability for the White reference group, consistent with systemic data-quality inequities in which some groups experience higher rates of recording problems.

Where possible, corruptions were instantiated using empirically observed transformations consistent with the sampled profile (e.g., common insertions, deletions and substitutions and their position patterns), rather than arbitrary perturbations, to preserve realism. For each record we stored whether a name was corrupted and the sampled corruption characteristics (edit distance, error type and position), enabling direct audit of both the exposure component (who was corrupted) and mechanism component (how corruption manifested).

## Stage 4: linkage models, calibration, and evaluation design

Each corrupted dataset was linked back to its uncorrupted counterpart, so that each record had a unique gold-standard match defined by the voter identifier. Each linkage model differed only in how forenames and surnames were represented, compared and weighted. This design isolates the contribution of name-comparator choices to linkage quality (Supplementary Figures 3, 4).

We compared three model families: (i) Jaro–Winkler or Levenshtein distance on forename and surname without TF adjustment, (ii) TF-adjusted Jaro–Winkler or Levenshtein distance on forename and surname, and (iii) a cluster-based embedding model in which forenames were compared using the embedding representation (Stage 2) combined with TF-adjusted Jaro–Winkler on surnames ("combined/cluster-based model"). All models used the same blocking and estimation strategy so that differences reflect name comparison functions and frequency handling (full blocking rules and model-fit diagnostics in Supplementary Materials). Linkages were performed using Splink (19), all codes are available on GitHub repository.

**Proportional stratified sampling for feasible evaluation**

Exhaustive evaluation at the candidate-pair level is computationally infeasible on the full extract, we constructed evaluation subsets using proportional stratified sampling within each corruption setting and replicate dataset. For each setting (1–3) and each replicate (v1–v5), we drew an evaluation sample comprising 5% of records, stratified jointly by ethnic group and corruption status. Corruption status was defined using indicators for whether the forename and/or surname had been corrupted (uncorrupted; forename-only; surname-only; both). Sampling fractions were proportional to each stratum's share in the full dataset under that setting/replicate, so that the evaluation subset preserved (i) the marginal ethnic-group distribution and (ii) the joint distribution of ethnic group and corruption exposure. This permits estimation of overall error rates and white-centric disparities without post-stratification weighting, while maintaining representation of corrupted records for subgroup comparisons. The evaluation of proportional sampling is reported in Supplementary Materials (Supplementary Table 1 and Figure 1).

**Outcomes, disparities, and statistical analysis**

We calibrated a match-weight threshold to target an overall missed-match rate closest to 0.20 across all settings and models. This was selected to reflect a pragmatic, label-scarce operating point that prioritises recall while limiting false matches to levels commonly used in administrative linkage pipelines. For each linkage run we classified each record as a correct match (linked to its gold record), a false match (linked to an incorrect record), or a missed match (not linked to its gold record). We report false match rates (FMR) and missed match rates (MMR) overall and stratified by ethnic groups.

We quantified inequality using white-centric disparities, defined as the difference in error rates between each ethnic group and Non-Hispanic White voters. Uncertainty intervals were derived from the five replicate corrupted datasets.

# Results

## Name characteristics and empirical error profiles

Across adjacent annual snapshot pairs, we observed 124,009 within-person forename discrepancies (0.17%) and 485,807 surname discrepancies (0.67%). These discrepancies provided empirical distributions of edit distance, edit type, and edit position, which were used to parameterise the synthetic corruption processes (Figure 1). In parallel, name-characteristic features varied across ethnic groups (Supplementary Table 2, 3), implying that linkage models relying on fixed comparator behaviour may operate under different between-person ambiguity regimes across populations.

Forenames among Asian voters were shorter on average and more commonly multi-token, whereas Non-Hispanic Black and Hispanic voters more often had longer single-token names and higher vowel–consonant ratios, consistent with distinct linguistic structures. These differences are consequential for comparator behaviour because a single edit represents a larger proportional perturbation in short strings, multi-term structure can shift how agreement is accumulated across name parts, and prefix-heavy similarity functions (e.g., Jaro–Winkler's upweighting of common prefixes) can behave differently when prefix distributions are population-specific.

## Overall linkage performance across models and settings

Table 1 reports overall false match rates and missed match rates at a fixed operating point: the match-weight threshold was calibrated in each setting to achieve an overall MMR closest to 20% in each model (Figure 2). Accordingly, differences in error rates across models should be interpreted as differences in how each model behaves under the same decision threshold, rather than as points chosen to optimise each model separately.

| Model | Corruption setting | FMR | MMR |
|---|---|---|---|
| jw | Setting 1 | 0.98% (0.98–0.99) | 19.97% (19.80–20.13) |
| jw | Setting 2 | 1.07% (1.05–1.10) | 20.45% (20.40–20.49) |
| jw | Setting 3 | 0.55% (0.54–0.57) | 20.34% (20.30–20.39) |
| levenshtein | Setting 1 | 1.02% (1.01–1.03) | 19.74% (19.70–19.79) |
| levenshtein | Setting 2 | 1.18% (1.15–1.22) | 19.90% (19.73–20.07) |
| levenshtein | Setting 3 | 0.56% (0.54–0.58) | 20.40% (20.37–20.43) |
| jw_no_tf | Setting 1 | 1.87% (1.80–1.95) | 19.13% (19.09–19.17) |
| jw_no_tf | Setting 2 | 2.03% (2.00–2.06) | 16.77% (16.75–16.80) |
| jw_no_tf | Setting 3 | 1.80% (1.76–1.85) | 10.84% (10.82–10.87) |
| levenshtein_no_tf | Setting 1 | 1.87% (1.80–1.95) | 19.13% (19.09–19.17) |
| levenshtein_no_tf | Setting 2 | 1.85% (1.70–2.00) | 19.04% (19.01–19.07) |
| levenshtein_no_tf | Setting 3 | 1.80% (1.77–1.82) | 10.84% (10.82–10.87) |
| combined | Setting 1 | 7.64% (7.59–7.69) | 18.08% (17.98–18.17) |
| combined | Setting 2 | 7.87% (7.73–8.01) | 17.63% (17.59–17.67) |
| combined | Setting 3 | 4.28% (4.22–4.35) | 18.96% (18.90–19.01) |

Table 1. Overall FMR and MMR by linkage model and corruption setting (mean; 95% CI across five replicate datasets). Jw: Jaro-Winkler Model, tf: Term-Frequency Adjustment.

Under Settings 1–2, the TF-adjusted Jaro–Winkler model achieved low FMR (Setting 1: 0.98% (0.98–1.00); Setting 2: 1.07% (1.05–1.10)) with MMR near the calibration target (Setting 1: 19.97% (19.86–20.12); Setting 2: 20.45% (20.40–20.49)). Removing TF adjustment increased false matching at the same operating point (Setting 1: 1.87% (1.82–1.97); Setting 2: 2.03% (2.01–2.05)), consistent with greater susceptibility to over-linkage among common names when distinctiveness is not incorporated into the evidence model.

Under Setting 3 (disproportionate exposure by ethnicity), TF-adjusted Jaro–Winkler retained very low overall FMR (0.55% (0.54–0.57)) while overall MMR remained close to the calibration target (20.34% (20.30–20.38)). The cluster-based embedding model exhibited a different pattern: MMR was slightly lower than TF-adjusted Jaro–Winkler (18.96% (18.91–19.00)) but with materially higher FMR (4.28% (4.20–4.32)) at a similar threshold. To characterise the full precision–recall behaviour beyond the calibrated operating point, Figure 3 presents MMR across the full range of match-weights for each model and setting.

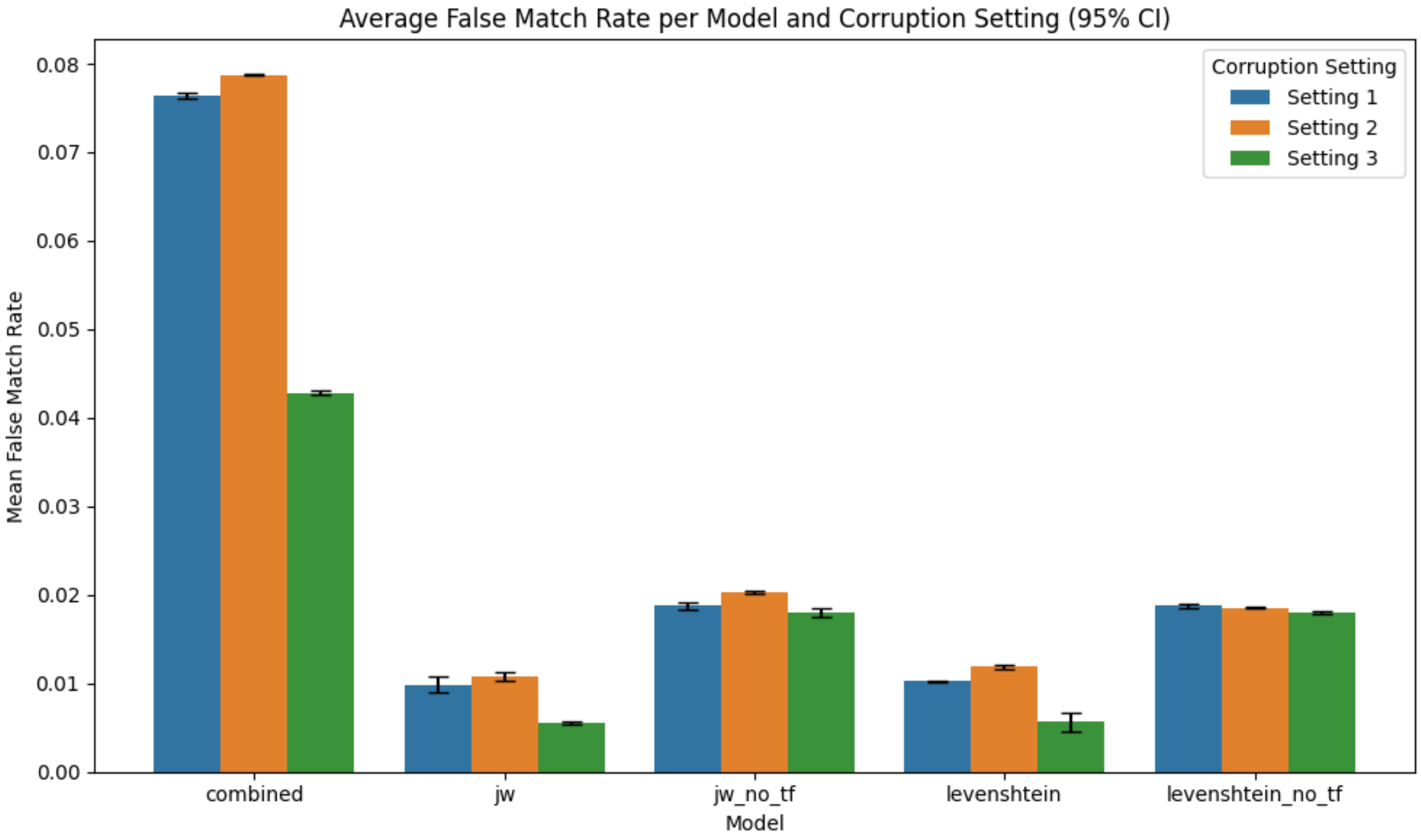

Figure 2. Overall false match rate by linkage model and corruption setting (mean; 95% CI across five replicates). Combined corresponds to cluster-based embeddings model. Jw: Jaro-Winkler Model, tf: Term-Frequency Adjustment.

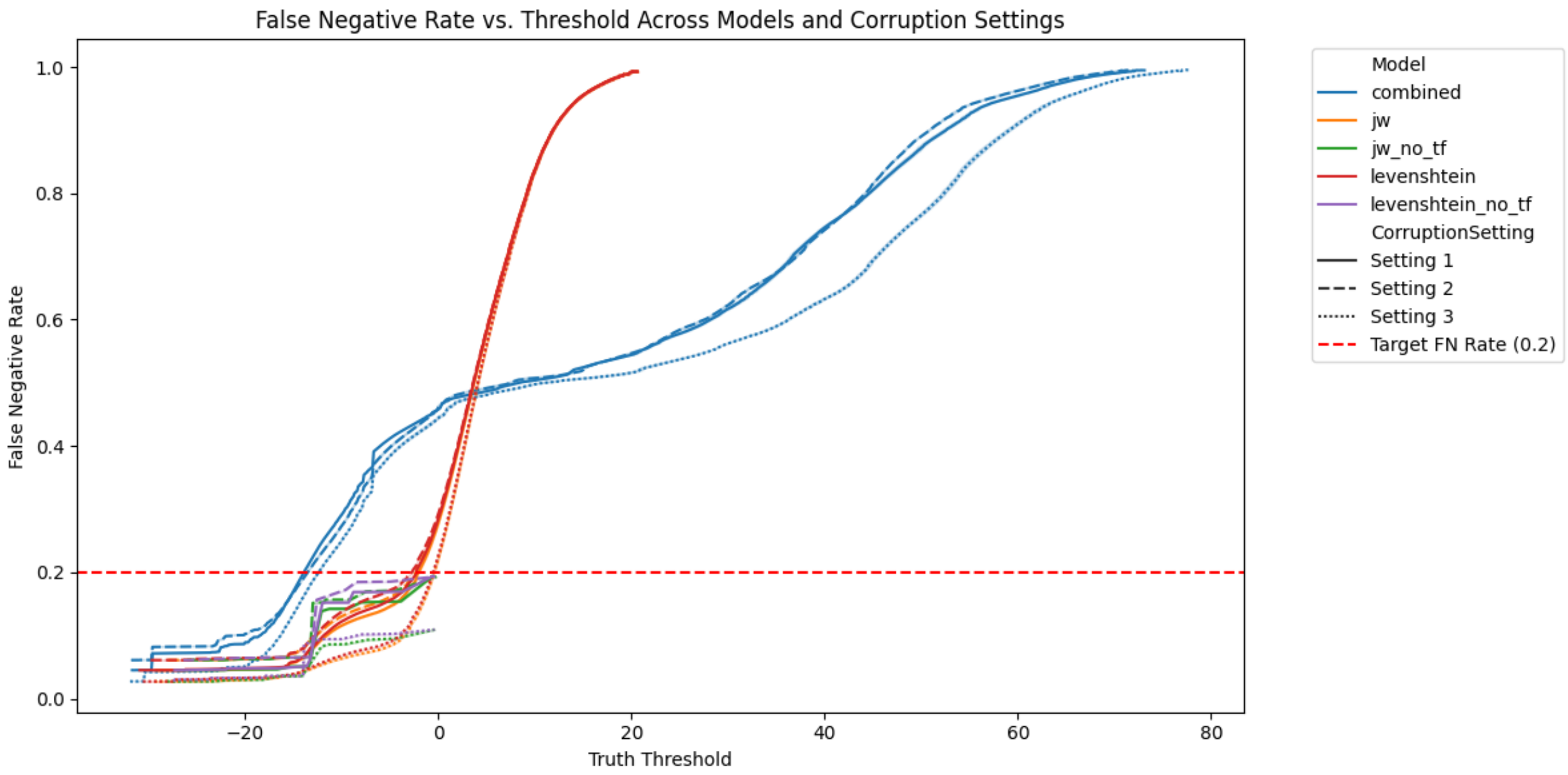

Figure 3. False negative rate (missed-match rate) versus match-weight threshold across linkage models and corruption settings. Jw: Jaro-Winkler Model, tf: Term-Frequency Adjustment, FN: False Negative.

## Equity impacts: reductions in white-centric missed-match disparities

Under Settings 1–2 (equal exposure), ethnic disparities were present but comparatively modest. In contrast, disproportionate exposure to corruption (Setting 3) amplified inequities: corruption rates ranged from approximately 1% among Non-Hispanic White voters to over one-third for forenames and over two-thirds for surnames among Indigenous or Pacific Islander voters in the simulated stress-test (Supplementary Figure 1).

Table 2 contrasts Setting 3 ethnic-group error rates and white-centric disparities across TF-adjusted Jaro–Winkler, its no-TF variant, and the cluster-based embedding model. TF-adjusted Jaro–Winkler produced large missed-match disparities for minoritised groups, most notably Indigenous or Pacific Islander voters (MMR 75.25% (75.06–75.48), disparity +60.0% (+59.8 to +60.2)) and Hispanic (White or Black) voters (MMR 51.52% (51.35–51.86), disparity +36.3% (+36.1 to +36.6)). These disparities persisted despite term-frequency adjustment, indicating that common-name penalties do not address the combination of high corruption exposure and structural mismatch (Figure 4).

| Model | Ethnic group | MMR | MMR disparity vs White (%) |
|---|---|---|---|
| jw_no_tf | Non-Hispanic White | 2.62% (2.59–2.64) | +0.0 (+0.0 to +0.0) |
| jw_no_tf | Non-Hispanic Black | 15.76% (15.67–15.84) | +13.1 (+13.0 to +13.2) |
| jw_no_tf | Hispanic (White or Black) | 63.05% (62.73–63.31) | +60.4 (+60.1 to +60.7) |
| jw_no_tf | Asian | 55.98% (55.57–56.52) | +53.4 (+53.0 to +53.9) |
| jw_no_tf | Indigenous or Pacific Islander | 77.84% (77.57–78.02) | +75.2 (+75.0 to +75.4) |
| jw | Non-Hispanic White | 15.25% (15.22–15.31) | +0.0 (+0.0 to +0.0) |
| jw | Non-Hispanic Black | 23.88% (23.80–23.99) | +8.6 (+8.6 to +8.7) |
| jw | Hispanic (White or Black) | 51.52% (51.35–51.86) | +36.3 (+36.1 to +36.6) |
| jw | Asian | 46.06% (45.78–46.43) | +30.8 (+30.5 to +31.2) |
| jw | Indigenous or Pacific Islander | 75.25% (75.06–75.48) | +60.0 (+59.8 to +60.2) |
| combined | Non-Hispanic White | 18.12% (18.08–18.18) | +0.0 (+0.0 to +0.0) |
| combined | Non-Hispanic Black | 18.72% (18.50–18.86) | +0.6 (+0.4 to +0.7) |
| combined | Hispanic (White or Black) | 28.28% (27.90–28.47) | +10.2 (+9.8 to +10.3) |
| combined | Asian | 21.92% (21.49–22.47) | +3.8 (+3.4 to +4.3) |
| combined | Indigenous or Pacific Islander | 34.43% (33.91–34.85) | +16.3 (+15.8 to +16.7) |

Table 2. Setting 3 ethnic-group missed match rates and white-centric disparities for no-TF Jaro–Winkler, TF-adjusted Jaro–Winkler and the cluster-based embedding model (mean; 95% CI across five replicates). Full results under all three settings are described in Supplementary Table 6. Jw: Jaro-Winkler Model, tf: Term-Frequency Adjustment.

The cluster-based embedding model substantially reduced missed-match disparities relative to TF-adjusted string similarity (e.g., Indigenous or Pacific Islander disparity +16.3% (+15.8 to +16.7); Hispanic disparity +10.2% (+9.8 to +10.3); Asian disparity +3.8% (+3.4 to +4.3)). However, this improvement in under-linkage equity came with higher false match rates, particularly in groups with high simulated corruption exposure, highlighting the need to jointly evaluate false and missed matches when selecting equity-oriented linkage specifications.

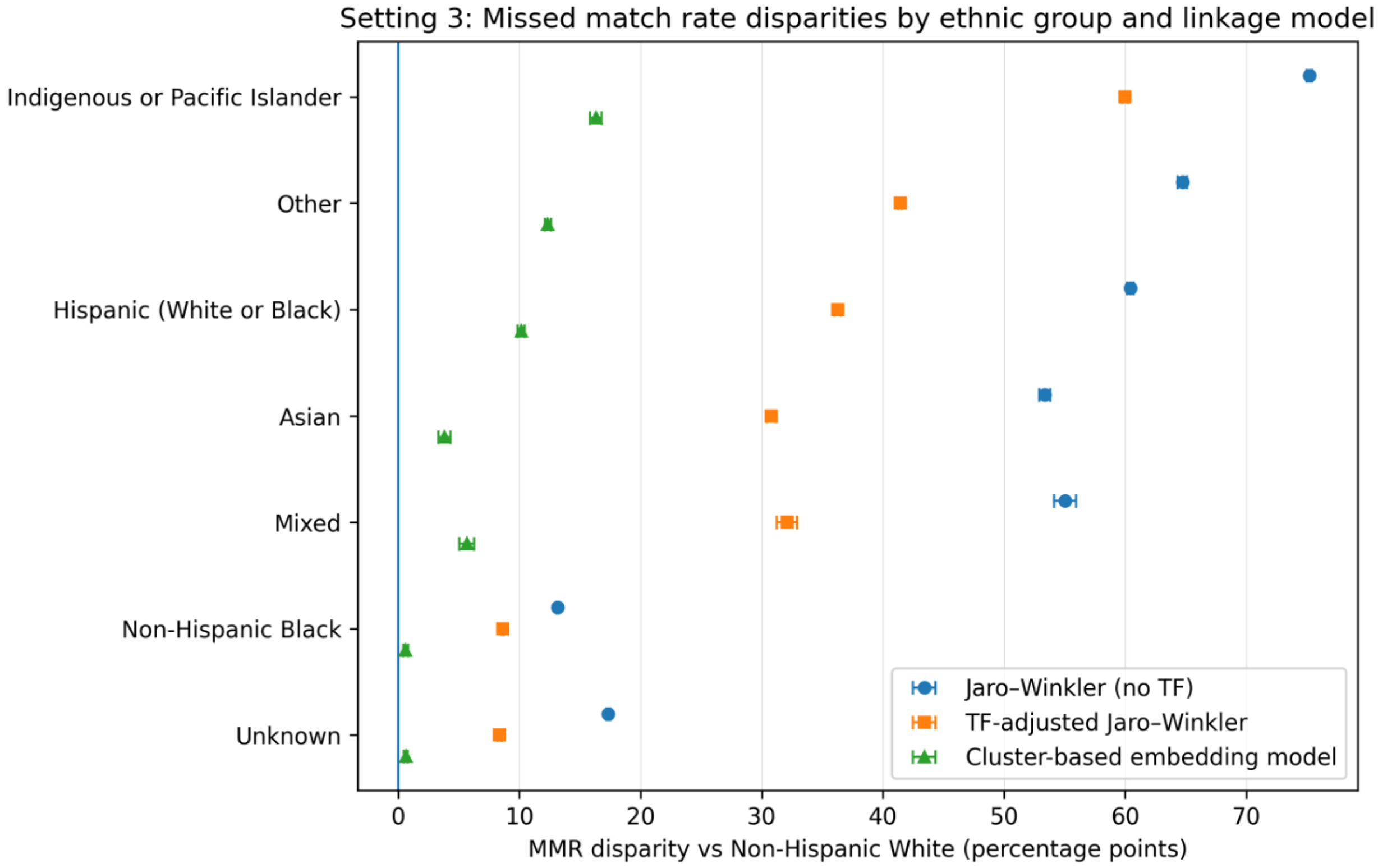

Figure 4. Setting 3 Missed Match Rate (MMR) disparity by ethnicity, comparing Jaro-Winkler (no TF), TF-adjusted Jaro-Winkler, and the cluster-based embedding model. Points show means across five replicates; error bars show 95% confidence intervals.

## Discussion

In this large administrative register, we show that ethnic disparities in name-based linkage are driven by the interaction of (i) heterogeneity in name structure, (ii) empirically realistic error mechanisms, and (iii) unequal exposure to error. Our main methodological contribution is a cluster-based embedding linkage approach that represents forenames via neighbourhoods in a low-dimensional name-characteristic space, providing an alternative to relying solely on string similarity and term-frequency adjustment.

The synthetic corruption settings clarify why error mechanism and exposure should both be considered in linkage equity assessments. When exposure was equal (Settings 1–2), errors differed less dramatically across groups even when error mechanisms differed. Under disproportionate exposure (Setting 3), missed-match disparities became extreme for some groups, consistent with a pathway whereby systemic data-quality inequities translate into differential under-linkage and biased epidemiologic inference.

This study confirms prior intuition in the literature on the use of TF-adjustment to reduce ethnic bias by providing the first direct, controlled evidence of its performance under ethnic-skewed corruption scenarios (3). Whilst TF adjustment offers a scalable, interpretable enhancement to string similarity models in equity-sensitive applications, it does not compensate for structural differences in how errors manifest across names or for large differences in exposure to those errors.

The cluster-based embedding model reduced missed-match disparities substantially compared with TF-adjusted Jaro–Winkler, particularly for groups with high simulated corruption exposure. The trade-off was increased false matches, reflecting that neighbourhood-based agreement can admit incorrect links when other discriminating information is absent. In applied linkage, this suggests that embedding-based comparisons should be paired with additional identifiers (e.g., date of birth, geography) or post-linkage constraints (e.g., one-to-one assignment, clerical review in high-ambiguity clusters) to improve equity without inflating over-linkage.

For epidemiologic users of linked administrative data, our findings reinforce that linkage quality is an equity-relevant measurement process (2,20). Studies should report subgroup-specific linkage error where feasible, and linkage pipelines should be stress-tested under plausible differential data-quality scenarios rather than assuming uniform error. Where linkage is performed centrally, data providers can mitigate inequities by improving upstream capture processes and by supporting linkage models that explicitly consider structure and exposure.

Our limitations include that the NCVR is a comparatively well-curated voter registry with very low observed within-person discrepancy rates; error mechanisms and may differ in settings where data capturing or entry is decentralised, multilingual naming is common, or data cleaning is minimal. Second, the ethnicity categories available in NCVR reflect optional self-reporting within a specific voter registry and may not map cleanly to other classification systems used in epidemiological studies, official statistics, or a different country (21). Many mechanisms relevant to name-based linkage, such as tokenisation and romanisation, relate more directly to linguistic and naming-system structure than to administrative ethnicity labels. Future work should evaluate whether linguistically informed groupings (e.g., by writing system, language family, or name-structure regimes) provide more actionable stratifications for identifying linkage vulnerability and targeting linkage model adaptations than broad ethnicity categories alone. Finally, linkage keys were deliberately restricted to names to isolate mechanisms of name-based bias. Real-world linkage systems often incorporate other identifiers such as date of birth. Excluding these features may understate model performance or bias mitigation capacity in practice, where additional contextual signals can potentially offset weaknesses in name comparisons.

## Conclusions

Ethnic disparities in name-based linkage are not an inherent property of names or of any single similarity metric; they arise from the interaction between model design, name structure, and unequal exposure to data quality problems. Cluster-based name embeddings offer a complementary tool to term-frequency adjustment for reducing under-linkage disparities, but must be deployed with explicit attention to precision trade-offs and to the structural sources of differential exposure.

## Ethics Approval

North Carolina has legislatively required the voter registry to be made public. The use of NCVR does not fall under the scope of UK GDPR, as it is publicly released under North Carolina state law. This project received research ethics approval from the UCL Ethics Committee (27715/001).

## Author Contributions

Conceptualization, Investigation, Statistical analysis, visualization, data access,verification and first draft: JL. Review and revision of manuscript: all authors. Supervision: MCB, RA, RB, KH.

## Supplementary Data

Supplementary data are available at *IJE* online.

## Conflict of interest

None declared.

## Funding

This work was supported by the Wellcome Trust [grant number 212953/Z/18/Z].

## Data Availability

The data underlying this article are available in North Carolina State Board of Elections repository. The datasets were derived from sources in the public domain: https://dl.ncsbe.gov/index.html?prefix=data/Snapshots/.

## Use of Artificial Intelligence (AI) tools

None declared.

## Figures Captions

Figure 1: Study design and Stage 1 empirical discrepancy profiling across annual North Carolina Voter Register snapshots (October 2011–October 2022).

Figure 2: Overall false match rate by linkage model and corruption setting (mean; 95% CI across five replicates). Combined corresponds to cluster-based embeddings model. Jw: Jaro-Winkler Model, tf: Term-Frequency Adjustment.

Figure 3. False negative rate (missed-match rate) versus match-weight threshold across linkage models and corruption settings. Jw: Jaro-Winkler Model, tf: Term-Frequency Adjustment, FN: False Negative.

Figure 4. Setting 3 Missed Match Rate (MMR) disparity by ethnicity, comparing Jaro-Winkler (no TF), TF-adjusted Jaro-Winkler, and the cluster-based embedding model. Points show means across five replicates; error bars show 95% confidence intervals.

# Supplementary Materials: Additional methods, diagnostics, and results for: Name-based linkage models and ethnic disparities in record linkage error

This document provides supplementary methodological detail and additional results supporting the main manuscript.

## Supplementary Methods

### Proportional stratified sampling design and sensitivity analysis

Where candidate-pair generation and scoring were computationally burdensome, we used proportional stratified sampling of records within each corruption setting and replicate dataset to construct evaluation subsets while preserving representativeness. Strata were defined by the cross-classification of ethnic group and corruption status (forename-only corrupted, surname-only corrupted, both corrupted, or uncorrupted). Records were sampled within each stratum at a constant fraction, so that the evaluation subset preserved (up to sampling variability) the marginal ethnic-group distribution and the joint distribution of ethnic group by corruption exposure. Achieved corruption exposure by ethnicity under Settings 1–3 is shown in Supplementary Figure 1. For Setting 3, the initial relative exposure weights and the adjusted (normalised) weights used in implementation are reported in Supplementary Table 1. Supplementary Table 2 describes distributional profile of error types, edit distances and positions by ethnicity, Supplementary Table 3 describes the absolute difference in forename feature distribution by ethnicity.

The group-specific corruption weights were designed to simulate plausible disparities in name error exposure, not to reflect precise empirical estimates. Higher weights for minoritised groups (e.g., Black, Hispanic, Other) reflect documented risks of mis-recording due to unfamiliarity, transliteration, or systemic bias. Asian names receive a moderate rate due to transliteration variability. White, Mixed, Indigenous, and Unknown are assigned the baseline corruption rate, either due to their statistical majority (and thus greater familiarity) or due to insufficient information to justify higher or lower risk. While Indigenous names may be vulnerable to specific issues (e.g., forced renaming, anglicisation), their small sample size and low cardinality complicate our corruption modelling. These weights are illustrative, intended to stress-test models under differential error exposure.

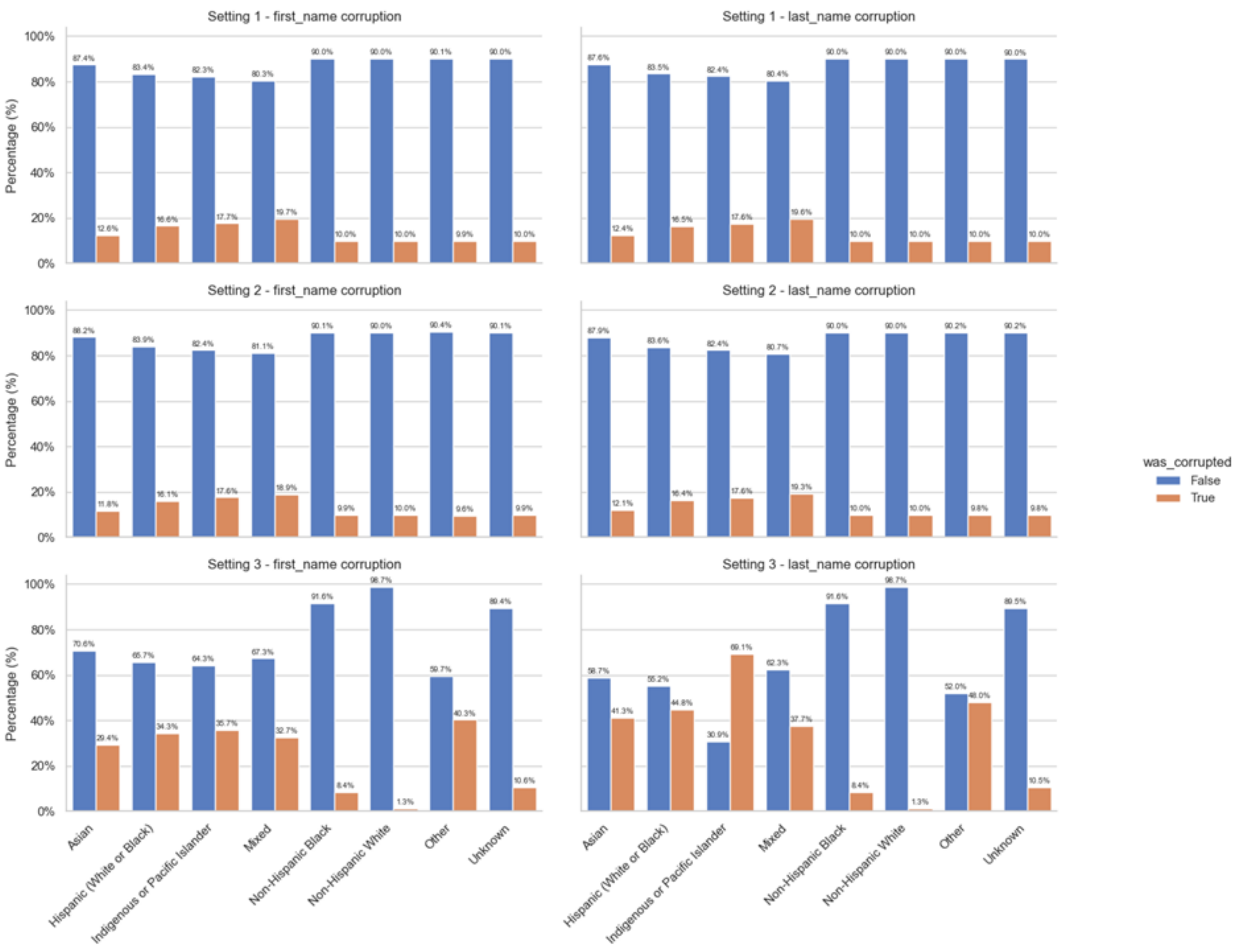

Supplementary Figure 1. Achieved corruption exposure by ethnicity under the three corruption settings, shown separately for forename and surname. Bars indicate the proportion of records corrupted (True) and not corrupted (False) within each ethnic group; Setting 1 targets uniform exposure, Setting 2 enforces equal exposure with ethnic-specific mechanisms, and Setting 3 imposes disproportionate exposure by design (see Methods).

| Ethnic group | Initial relative weight | Adjusted weight used |
|---|---|---|
| Non-Hispanic White | 0.1 | 0.087 |
| Non-Hispanic Black | 0.2 | 0.1739 |
| Hispanic (White or Black) | 0.2 | 0.0836 |
| Asian | 0.15 | 0.1373 |
| Other | 0.2 | 0.2972 |
| Mixed | 0.1 | 0.0603 |
| Indigenous or Pacific Islander | 0.1 | 0.0738 |
| Unknown | 0.1 | 0.087 |

Supplementary Table 1. Setting 3 exposure weights by ethnic group. Initial relative weights were specified a priori to allocate the fixed Setting 3 corruption budget across ethnic groups. Corruption was implemented by allocating an exact number of corruptions (10% overall) rather than by independent Bernoulli draws. We therefore (1) rescaled the initial weights to sum to 1, (2) converted them to integer target counts per group, and (3) reconciled rounding and eligibility constraints (e.g., missing or non-corruptible names) by redistributing any unallocated corruptions to groups with remaining capacity in proportion to their weights. The adjusted weights here correspond to the realised shares of the corruption budget assigned to each group.

| Ethnic group | del | ins | rep | 1 | 2 | 3 | 4 | 5 | 6 | 7+ | Across | End | First Half | Secd Half | Start |
|---|---|---|---|---|---|---|---|---|---|---|---|---|---|---|---|
| Asian | 38.6% | 41.1% | 20.3% | 8.1% | 4.3% | 7.2% | 17.1% | 20.6% | 15.9% | 26.7% | 41.9% | 16.6% | 2.9% | 38.4% | 0.3% |
| Hispanic (White or Black) | 37.0% | 36.5% | 26.5% | 11.3% | 10.1% | 4.4% | 4.8% | 12.0% | 22.2% | 35.2% | 51.8% | 9.1% | 2.7% | 35.8% | 0.5% |
| Indigenous or Pacific Islander | 35.4% | 34.4% | 30.3% | 13.3% | 17.8% | 8.0% | 10.1% | 16.2% | 12.3% | 22.3% | 46.3% | 13.6% | 2.0% | 37.8% | 0.3% |
| Mixed | 34.7% | 36.2% | 29.1% | 18.2% | 13.6% | 5.2% | 6.9% | 10.9% | 16.6% | 28.4% | 42.1% | 11.3% | 4.9% | 41.2% | 0.5% |
| Non-Hispanic Black | 34.7% | 34.1% | 31.2% | 22.2% | 20.6% | 6.4% | 7.3% | 10.6% | 13.9% | 19.1% | 40.2% | 11.6% | 5.5% | 42.3% | 0.4% |
| Non-Hispanic White | 31.9% | 36.3% | 31.8% | 6.4% | 8.6% | 7.3% | 12.4% | 19.3% | 20.7% | 25.3% | 58.6% | 9.7% | 1.7% | 30.0% | 0.1% |
| Other | 35.8% | 38.2% | 26.0% | 10.7% | 6.7% | 5.3% | 11.8% | 15.9% | 18.1% | 31.5% | 49.0% | 11.6% | 3.1% | 35.9% | 0.3% |
| Unknown | 33.6% | 35.8% | 30.6% | 13.0% | 12.2% | 7.0% | 9.0% | 14.9% | 17.1% | 26.7% | 48.4% | 10.2% | 3.9% | 37.0% | 0.5% |

| Ethnic group | del | ins | rep | 1 | 2 | 3 | 4 | 5 | 6 | 7+ | Across | End | First Half | Secd Half | Start |
|---|---|---|---|---|---|---|---|---|---|---|---|---|---|---|---|
| Asian | 26.2% | 30.7% | 43.1% | 1.9% | 1.3% | 2.7% | 8.0% | 15.2% | 20.0% | 50.9% | 87.2% | 1.0% | 3.1% | 8.5% | 0.2% |
| Hispanic (White or Black) | 31.4% | 34.8% | 33.8% | 1.5% | 0.5% | 0.5% | 2.6% | 11.1% | 20.2% | 63.5% | 77.2% | 3.1% | 8.5% | 11.1% | 0.0% |
| Indigenous or Pacific Islander | 28.8% | 29.0% | 42.2% | 0.4% | 0.3% | 0.3% | 3.3% | 12.4% | 22.9% | 60.4% | 92.9% | 0.5% | 2.0% | 4.6% | 0.0% |
| Mixed | 30.4% | 32.2% | 37.4% | 1.9% | 0.6% | 0.9% | 3.5% | 12.2% | 20.8% | 60.0% | 83.7% | 2.2% | 6.1% | 8.0% | 0.1% |
| Non-Hispanic Black | 28.3% | 30.3% | 41.4% | 0.7% | 0.3% | 0.7% | 4.4% | 14.7% | 25.5% | 53.6% | 90.1% | 1.0% | 2.9% | 6.0% | 0.0% |
| Non-Hispanic White | 28.2% | 27.3% | 44.5% | 0.4% | 0.2% | 0.7% | 4.4% | 14.9% | 25.5% | 54.0% | 96.2% | 0.3% | 1.1% | 2.3% | 0.0% |
| Other | 30.5% | 34.2% | 35.3% | 1.2% | 0.4% | 0.7% | 2.8% | 10.6% | 20.1% | 64.2% | 80.1% | 2.2% | 8.3% | 9.3% | 0.0% |
| Unknown | 30.0% | 32.4% | 37.6% | 2.1% | 0.90% | 0.8% | 3.6% | 12.3% | 22.1% | 58.2% | 82.7% | 1.8% | 6.3% | 9.2% | 0.1% |

Supplementary Table 2. Forename (top) and surname (bottom) error profile describing distribution of error types, edit distances and positions of error, derived empirically from NCVR 2011-2022 Annual snapshots record pairs. Del: Deletion, Ins: Insertion, Rep: Replacement.

| first names (mean, sd) | Asian | Hispanic (White or Black) | Indigenous or Pacific Islander | Mixed | Non-Hispanic Black | Non-Hispanic White | Other | Unknown |
|---|---|---|---|---|---|---|---|---|
| length without space | 5.89 (2.03) | 6.17 (1.56) | 6.03 (1.46) | 6.23 (1.52) | 6.28 (1.43) | 5.95 (1.46) | 6.09 (1.64) | 6.11 (1.53) |
| length with space | 5.93 (2.1) | 6.17 (1.57) | 6.03 (1.47) | 6.23 (1.53) | 6.28 (1.44) | 5.95 (1.46) | 6.1 (1.66) | 6.12 (1.54) |
| number of terms | 1.04 (0.2) | 1.01 (0.08) | 1 (0.06) | 1 (0.06) | 1 (0.05) | 1 (0.05) | 1.01 (0.09) | 1 (0.07) |
| character per term | 5.72 (1.92) | 6.14 (1.53) | 6.01 (1.45) | 6.21 (1.5) | 6.27 (1.43) | 5.94 (1.45) | 6.06 (1.6) | 6.09 (1.51) |
| vowel consonant ratio | 0.77 (0.43) | 0.85 (0.41) | 0.7 (0.37) | 0.78 (0.4) | 0.75 (0.38) | 0.67 (0.34) | 0.83 (0.4) | 0.73 (0.38) |
| number of names with shared first three letters | 25128.23 (47169.62) | 49654.92 (69075.12) | 47375.02 (58482.2) | 45383.45 (57637.09) | 44349.29 (58371.96) | 57959.05 (60566.75) | 41695.26 (63389.28) | 48703.71 (59488.87) |

Supplementary Table 3. Absolute difference in forename feature distributions by ethnicity. Sd: standard deviations.

## Cluster-based name-embedding model specification and robustness

The cluster-based model operationalises forename similarity using proximity in a low-dimensional embedding of name-structure features. Name features were standardised and summarised via principal components analysis. Candidate pairs were compared using absolute differences on the first eight principal components (Supplementary Table 4). Differences were discretised into ordinal comparison levels using empirical percentile thresholds derived from within-person discrepancy distributions (e.g., <5th, 5–10th, 10–25th, 25–50th, >50th percentiles), yielding a neighbourhood-based notion of agreement, which is used as levels of comparison in probabilistic linkage model (Supplementary Figure 2, 3).

| Feature | PC1 | PC2 | PC3 | PC4 | PC5 | PC6 | PC7 | PC8 |
|---|---|---|---|---|---|---|---|---|
| Hyphen | 0.085 | 0.021 | -0.113 | 0.286 | 0.061 | 0.845 | 0.233 | 0.219 |
| Apostrophe | 0.040 | 0.035 | -0.127 | 0.284 | 0.774 | -0.225 | -0.353 | 0.220 |
| Length (with space) | 0.563 | -0.067 | -0.010 | -0.079 | -0.023 | -0.046 | 0.012 | 0.014 |
| Length (without space) | 0.563 | -0.065 | -0.015 | -0.058 | -0.038 | -0.063 | 0.025 | 0.023 |
| Vowel-Consonant ratio | -0.001 | 0.547 | 0.371 | 0.048 | 0.002 | 0.011 | -0.003 | -0.014 |
| Longest vowel | 0.127 | 0.386 | 0.437 | 0.124 | 0.033 | 0.020 | -0.012 | -0.482 |
| Unique binary | 0.141 | 0.133 | -0.253 | 0.563 | 0.129 | 0.030 | 0.095 | -0.369 |
| Uniqueness (continuous) | -0.020 | -0.400 | 0.499 | 0.238 | 0.024 | 0.030 | -0.070 | -0.057 |
| Number of terms | 0.064 | 0.043 | -0.134 | 0.539 | -0.401 | -0.451 | 0.334 | 0.237 |
| Middle name (binary) | -0.046 | -0.097 | 0.147 | -0.201 | 0.463 | -0.146 | 0.829 | -0.038 |
| Shared trigram | 0.031 | -0.344 | 0.520 | 0.260 | -0.026 | 0.016 | -0.064 | 0.297 |
| Ends with vowel | 0.083 | 0.478 | 0.139 | -0.093 | 0.006 | -0.021 | 0.035 | 0.620 |
| Characters per term | 0.554 | -0.074 | 0.012 | -0.168 | 0.042 | 0.028 | -0.043 | -0.024 |

Supplementary Table 4. PCA loadings for forename embedding components (PC1–PC8). Loadings were estimated on the October 2022 uncorrupted extract. Component signs are arbitrary; interpretation should focus on the relative magnitude and pattern of loadings within each component. Loadings are rounded to three decimal places.

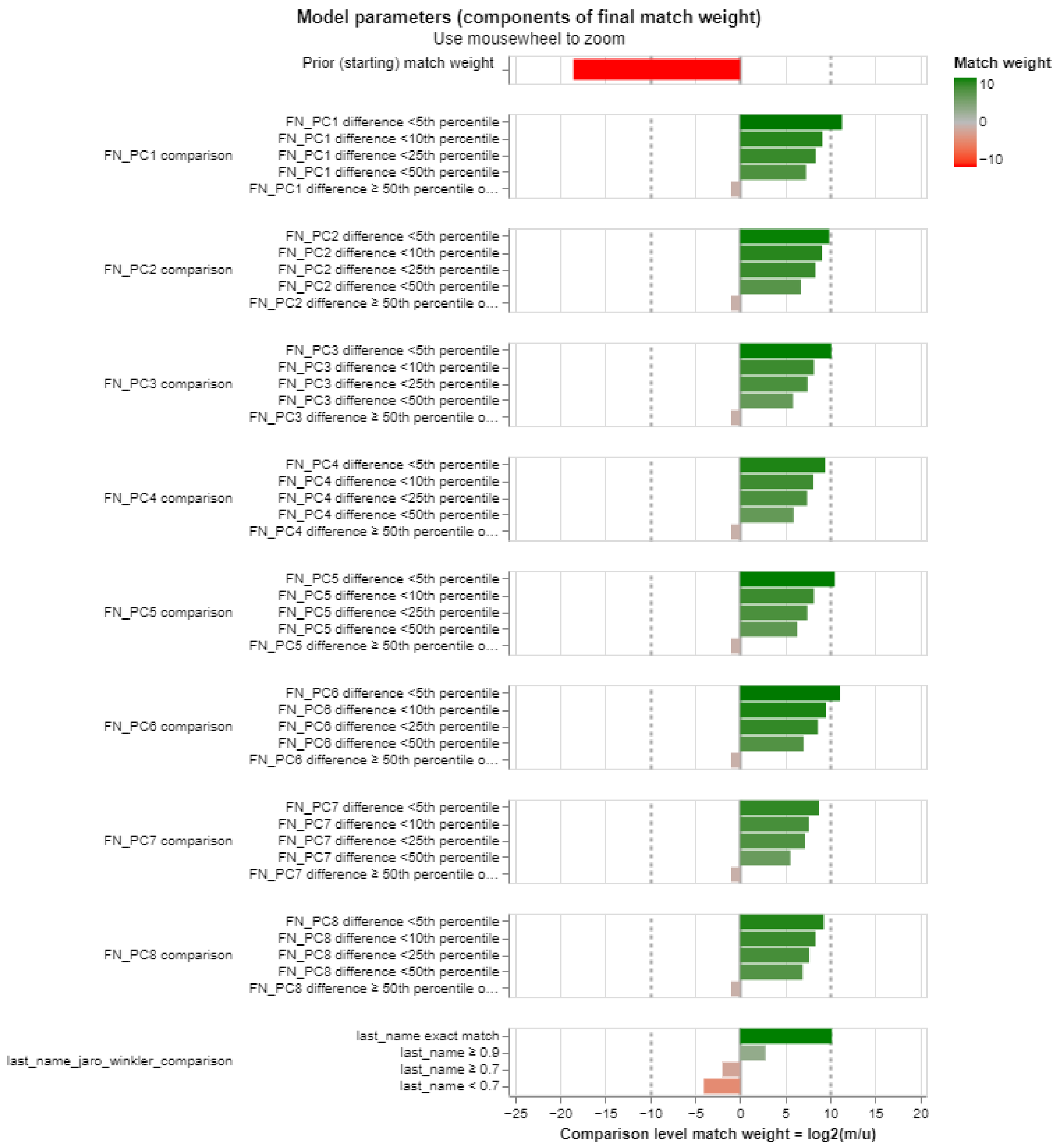

Supplementary Figure 2. Estimated comparison weights for the cluster-based model (forename PCA-difference comparisons and surname comparator).

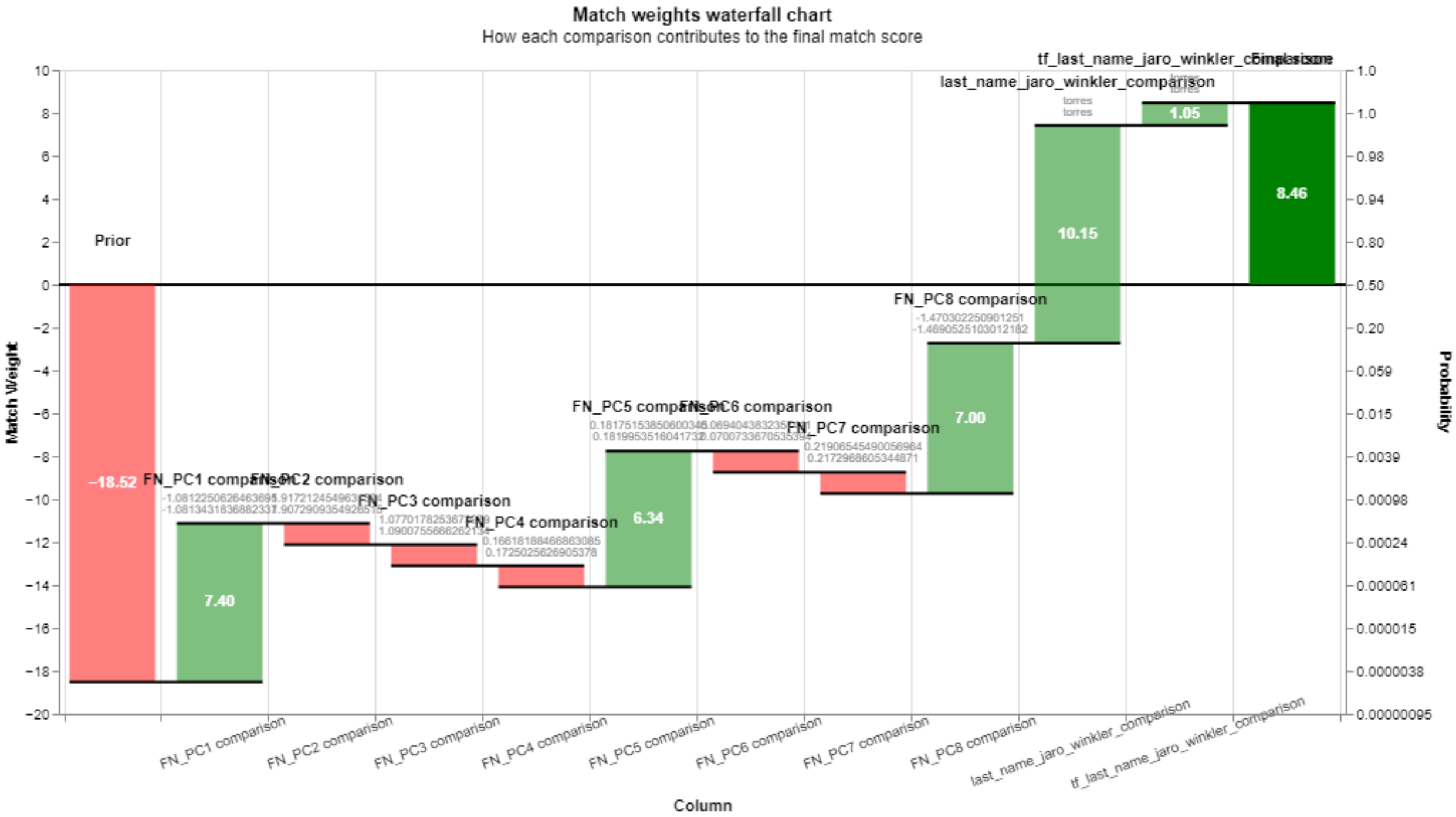

Supplementary Figure 3. Waterfall chart illustrating how comparison components contribute to the final match weight under the cluster-based model. Figure is showing a comparison of example record pair "Aaron Torres" with "Aaron Torres" using the feature-based embedding model.

A waterfall chart visualises how the overall match weight for a record pair is built up from the prior weight and the contribution of each comparison field. Each bar represents the positive or negative weight added by a specific comparison, and the final bar shows the cumulative match weight, which is then converted into a match probability. The small grey numbers represent the normalised PC loadings for each of the linked dataset.

## Blocking rules and its limitations

We applied two blocking rules across all models: (1) exact agreement on birth year, gender, and last name combined with agreement on the first three characters of the first name, and (2) agreement on birth year and gender with agreement on the first three characters of the forename. This design deliberately balances between reducing the candidate space and preserving plausible matches, especially under conditions of name corruption, while maintaining a computationally feasible number of comparison pairs. While necessary for scalability, this blocking strategy restricts the evaluation to name pairs with minimal prefix corruption. This has two implications. First, it inherently favours string similarity algorithms like Jaro-Winkler, which heavily weight early-character similarity. As a result, their performance may appear artificially stronger, particularly in terms of recall, since corrupted prefix cases, where Jaro-Winkler would typically fail, are excluded from consideration. Similarly, it favours Levenstein comparison as the maximum possible space for transposition is restricted. In effect, the current approach likely underestimates the true missed match rate for string-based models, especially in high-corruption scenarios. Therefore, the demonstrated performance and equity benefits of the combined model are likely conservative. In less restrictive linkage environments, where more complex or prefix-based corruptions are included, the relative advantage of feature-based or hybrid models may be even greater in terms of linkage equity, but likely at a cost of higher false matches.

## Supplementary Results

Supplementary Table 5, 6 provide extended quantitative results across all evaluated models (including Levenshtein-based comparators) and across all corruption settings. Rates are expressed as percentages, and disparities are expressed in percentage points relative to Non-Hispanic White voters.

| Model | CorruptionSetting | ethnic_group | MMR disparity vs White (pp, 95% CI) |
|---|---|---|---|
| combined | Setting 1 | Asian | -10.4 (-10.7 to -10.2) |
| combined | Setting 1 | Hispanic (White or Black) | -3.6 (-3.9 to -3.2) |
| combined | Setting 1 | Indigenous or Pacific Islander | +5.5 (+5.1 to +5.8) |
| combined | Setting 1 | Mixed | +0.3 (-0.4 to +1.0) |
| combined | Setting 1 | Non-Hispanic Black | -2.9 (-3.2 to -2.5) |
| combined | Setting 1 | Non-Hispanic White | +0.0 (+0.0 to +0.0) |
| combined | Setting 1 | Other | -9.4 (-9.6 to -9.2) |
| combined | Setting 1 | Unknown | -3.5 (-3.7 to -3.3) |
| jw | Setting 1 | Asian | -1.3 (-1.6 to -1.0) |
| jw | Setting 1 | Hispanic (White or Black) | +5.2 (+5.0 to +5.4) |
| jw | Setting 1 | Indigenous or Pacific Islander | +11.8 (+11.3 to +12.4) |
| jw | Setting 1 | Mixed | +11.4 (+10.9 to +11.9) |
| jw | Setting 1 | Non-Hispanic Black | +0.1 (+0.1 to +0.2) |
| jw | Setting 1 | Non-Hispanic White | +0.0 (+0.0 to +0.0) |
| jw | Setting 1 | Other | -4.5 (-4.8 to -4.3) |
| jw | Setting 1 | Unknown | -1.6 (-1.8 to -1.4) |
| jw_no_tf | Setting 1 | Asian | +4.4 (+4.0 to +4.8) |
| jw_no_tf | Setting 1 | Hispanic (White or Black) | +11.9 (+11.7 to +12.2) |
| jw_no_tf | Setting 1 | Indigenous or Pacific Islander | +13.6 (+13.4 to +13.9) |
| jw_no_tf | Setting 1 | Mixed | +17.4 (+17.0 to +17.7) |
| jw_no_tf | Setting 1 | Non-Hispanic Black | +0.1 (+0.0 to +0.2) |
| jw_no_tf | Setting 1 | Non-Hispanic White | +0.0 (+0.0 to +0.0) |
| jw_no_tf | Setting 1 | Other | +0.0 (-0.2 to +0.2) |
| jw_no_tf | Setting 1 | Unknown | +0.2 (+0.1 to +0.4) |
| levenshtein | Setting 1 | Asian | -2.8 (-3.2 to -2.5) |
| levenshtein | Setting 1 | Hispanic (White or Black) | +5.5 (+5.2 to +5.9) |
| levenshtein | Setting 1 | Indigenous or Pacific Islander | +11.4 (+11.0 to +11.9) |
| levenshtein | Setting 1 | Mixed | +11.3 (+10.9 to +11.7) |
| levenshtein | Setting 1 | Non-Hispanic Black | -0.9 (-0.9 to -0.8) |
| levenshtein | Setting 1 | Non-Hispanic White | +0.0 (+0.0 to +0.0) |
| levenshtein | Setting 1 | Other | -4.6 (-4.9 to -4.4) |
| levenshtein | Setting 1 | Unknown | -1.9 (-2.1 to -1.7) |
| levenshtein_no_tf | Setting 1 | Asian | +4.4 (+4.0 to +4.8) |
| levenshtein_no_tf | Setting 1 | Hispanic (White or Black) | +11.9 (+11.7 to +12.2) |
| levenshtein_no_tf | Setting 1 | Indigenous or Pacific Islander | +13.6 (+13.4 to +13.9) |
| levenshtein_no_tf | Setting 1 | Mixed | +17.4 (+17.0 to +17.7) |
| levenshtein_no_tf | Setting 1 | Non-Hispanic Black | +0.1 (+0.0 to +0.2) |
| levenshtein_no_tf | Setting 1 | Non-Hispanic White | +0.0 (+0.0 to +0.0) |
| levenshtein_no_tf | Setting 1 | Other | +0.0 (-0.2 to +0.2) |
| levenshtein_no_tf | Setting 1 | Unknown | +0.2 (+0.1 to +0.4) |
| combined | Setting 2 | Asian | -9.2 (-9.5 to -8.9) |
| combined | Setting 2 | Hispanic (White or Black) | -3.3 (-3.8 to -2.9) |
| combined | Setting 2 | Indigenous or Pacific Islander | +3.9 (+3.5 to +4.4) |
| combined | Setting 2 | Mixed | -0.8 (-1.5 to -0.1) |
| combined | Setting 2 | Non-Hispanic Black | -3.5 (-3.7 to -3.3) |
| combined | Setting 2 | Non-Hispanic White | +0.0 (+0.0 to +0.0) |
| combined | Setting 2 | Other | -8.8 (-9.1 to -8.6) |
| combined | Setting 2 | Unknown | -4.2 (-4.3 to -4.0) |
| jw | Setting 2 | Asian | -2.5 (-2.8 to -2.2) |
| jw | Setting 2 | Hispanic (White or Black) | +4.8 (+4.6 to +5.0) |
| jw | Setting 2 | Indigenous or Pacific Islander | +11.3 (+10.9 to +11.7) |
| jw | Setting 2 | Mixed | +10.5 (+10.3 to +10.7) |
| jw | Setting 2 | Non-Hispanic Black | -1.1 (-1.2 to -1.0) |
| jw | Setting 2 | Non-Hispanic White | +0.0 (+0.0 to +0.0) |
| jw | Setting 2 | Other | -5.0 (-5.2 to -4.7) |
| jw | Setting 2 | Unknown | -2.5 (-2.5 to -2.5) |
| jw_no_tf | Setting 2 | Asian | +3.3 (+2.8 to +3.7) |
| jw_no_tf | Setting 2 | Hispanic (White or Black) | +10.2 (+9.9 to +10.5) |
| jw_no_tf | Setting 2 | Indigenous or Pacific Islander | +11.0 (+10.6 to +11.3) |
| jw_no_tf | Setting 2 | Mixed | +13.0 (+12.6 to +13.3) |
| jw_no_tf | Setting 2 | Non-Hispanic Black | -1.4 (-1.4 to -1.4) |
| jw_no_tf | Setting 2 | Non-Hispanic White | +0.0 (+0.0 to +0.0) |
| jw_no_tf | Setting 2 | Other | -0.2 (-0.4 to -0.1) |
| jw_no_tf | Setting 2 | Unknown | -0.3 (-0.4 to -0.2) |
| levenshtein | Setting 2 | Asian | -2.7 (-3.1 to -2.4) |
| levenshtein | Setting 2 | Hispanic (White or Black) | +6.3 (+6.1 to +6.6) |
| levenshtein | Setting 2 | Indigenous or Pacific Islander | +10.9 (+10.4 to +11.4) |
| levenshtein | Setting 2 | Mixed | +11.1 (+10.7 to +11.6) |
| levenshtein | Setting 2 | Non-Hispanic Black | -1.9 (-2.0 to -1.7) |
| levenshtein | Setting 2 | Non-Hispanic White | +0.0 (+0.0 to +0.0) |
| levenshtein | Setting 2 | Other | -4.1 (-4.6 to -3.7) |
| levenshtein | Setting 2 | Unknown | -2.4 (-2.4 to -2.4) |
| levenshtein_no_tf | Setting 2 | Asian | +3.6 (+3.2 to +4.1) |
| levenshtein_no_tf | Setting 2 | Hispanic (White or Black) | +11.4 (+11.1 to +11.8) |
| levenshtein_no_tf | Setting 2 | Indigenous or Pacific Islander | +13.4 (+12.9 to +13.9) |
| levenshtein_no_tf | Setting 2 | Mixed | +16.7 (+16.5 to +17.0) |
| levenshtein_no_tf | Setting 2 | Non-Hispanic Black | -0.1 (-0.1 to -0.0) |
| levenshtein_no_tf | Setting 2 | Non-Hispanic White | +0.0 (+0.0 to +0.0) |
| levenshtein_no_tf | Setting 2 | Other | -0.5 (-0.7 to -0.2) |
| levenshtein_no_tf | Setting 2 | Unknown | -0.0 (-0.1 to +0.0) |
| combined | Setting 3 | Asian | +3.8 (+3.3 to +4.3) |
| combined | Setting 3 | Hispanic (White or Black) | +10.2 (+9.9 to +10.5) |
| combined | Setting 3 | Indigenous or Pacific Islander | +16.3 (+15.8 to +16.8) |
| combined | Setting 3 | Mixed | +5.7 (+5.0 to +6.3) |
| combined | Setting 3 | Non-Hispanic Black | +0.6 (+0.4 to +0.8) |
| combined | Setting 3 | Non-Hispanic White | +0.0 (+0.0 to +0.0) |
| combined | Setting 3 | Other | +12.4 (+12.1 to +12.6) |
| combined | Setting 3 | Unknown | +0.6 (+0.5 to +0.8) |
| jw | Setting 3 | Asian | +30.8 (+30.4 to +31.2) |
| jw | Setting 3 | Hispanic (White or Black) | +36.3 (+36.0 to +36.5) |
| jw | Setting 3 | Indigenous or Pacific Islander | +60.0 (+59.8 to +60.2) |
| jw | Setting 3 | Mixed | +32.1 (+31.2 to +32.9) |
| jw | Setting 3 | Non-Hispanic Black | +8.6 (+8.5 to +8.7) |
| jw | Setting 3 | Non-Hispanic White | +0.0 (+0.0 to +0.0) |
| jw | Setting 3 | Other | +41.4 (+41.2 to +41.6) |
| jw | Setting 3 | Unknown | +8.4 (+8.2 to +8.5) |
| jw_no_tf | Setting 3 | Asian | +53.4 (+52.9 to +53.8) |
| jw_no_tf | Setting 3 | Hispanic (White or Black) | +60.4 (+60.2 to +60.7) |
| jw_no_tf | Setting 3 | Indigenous or Pacific Islander | +75.2 (+75.0 to +75.5) |
| jw_no_tf | Setting 3 | Mixed | +55.0 (+54.1 to +56.0) |
| jw_no_tf | Setting 3 | Non-Hispanic Black | +13.1 (+13.0 to +13.2) |
| jw_no_tf | Setting 3 | Non-Hispanic White | +0.0 (+0.0 to +0.0) |
| jw_no_tf | Setting 3 | Other | +64.7 (+64.3 to +65.1) |
| jw_no_tf | Setting 3 | Unknown | +17.3 (+17.2 to +17.5) |
| levenshtein | Setting 3 | Asian | +35.4 (+34.9 to +35.8) |
| levenshtein | Setting 3 | Hispanic (White or Black) | +42.7 (+42.4 to +43.1) |
| levenshtein | Setting 3 | Indigenous or Pacific Islander | +61.7 (+61.4 to +62.0) |
| levenshtein | Setting 3 | Mixed | +35.8 (+35.1 to +36.5) |
| levenshtein | Setting 3 | Non-Hispanic Black | +9.2 (+9.1 to +9.2) |
| levenshtein | Setting 3 | Non-Hispanic White | +0.0 (+0.0 to +0.0) |
| levenshtein | Setting 3 | Other | +47.4 (+47.2 to +47.6) |
| levenshtein | Setting 3 | Unknown | +9.5 (+9.3 to +9.7) |
| levenshtein_no_tf | Setting 3 | Asian | +53.4 (+52.9 to +53.8) |
| levenshtein_no_tf | Setting 3 | Hispanic (White or Black) | +60.4 (+60.2 to +60.7) |
| levenshtein_no_tf | Setting 3 | Indigenous or Pacific Islander | +75.2 (+75.0 to +75.5) |
| levenshtein_no_tf | Setting 3 | Mixed | +55.0 (+54.1 to +56.0) |
| levenshtein_no_tf | Setting 3 | Non-Hispanic Black | +13.1 (+13.0 to +13.2) |
| levenshtein_no_tf | Setting 3 | Non-Hispanic White | +0.0 (+0.0 to +0.0) |
| levenshtein_no_tf | Setting 3 | Other | +64.7 (+64.3 to +65.1) |
| levenshtein_no_tf | Setting 3 | Unknown | +17.3 (+17.2 to +17.5) |

Supplementary Table 5. White-centric disparities in missed match rate (MMR) by model, setting, and ethnic group (percentage points; mean and 95% CI across five replicates).

| Model | Corruption Setting | ethnic_group | FMR disparity vs White (pp, 95% CI) |
|---|---|---|---|
| combined | Setting 1 | Asian | -3.1 (-3.4 to -2.9) |
| combined | Setting 1 | Hispanic (White or Black) | +2.0 (+1.8 to +2.3) |
| combined | Setting 1 | Indigenous or Pacific Islander | +5.2 (+4.9 to +5.4) |
| combined | Setting 1 | Mixed | +5.8 (+5.4 to +6.3) |
| combined | Setting 1 | Non-Hispanic Black | -0.2 (-0.3 to -0.2) |
| combined | Setting 1 | Non-Hispanic White | +0.0 (+0.0 to +0.0) |
| combined | Setting 1 | Other | -2.8 (-3.0 to -2.6) |
| combined | Setting 1 | Unknown | -4.0 (-4.1 to -3.9) |
| jw | Setting 1 | Asian | -0.7 (-0.8 to -0.6) |
| jw | Setting 1 | Hispanic (White or Black) | -0.2 (-0.4 to +0.1) |
| jw | Setting 1 | Indigenous or Pacific Islander | +0.4 (+0.2 to +0.6) |
| jw | Setting 1 | Mixed | +0.3 (+0.1 to +0.5) |
| jw | Setting 1 | Non-Hispanic Black | +0.1 (+0.0 to +0.1) |
| jw | Setting 1 | Non-Hispanic White | +0.0 (+0.0 to +0.0) |
| jw | Setting 1 | Other | -0.6 (-0.6 to -0.5) |
| jw | Setting 1 | Unknown | -0.6 (-0.6 to -0.5) |
| jw_no_tf | Setting 1 | Asian | -1.6 (-1.8 to -1.5) |
| jw_no_tf | Setting 1 | Hispanic (White or Black) | -0.9 (-1.0 to -0.7) |
| jw_no_tf | Setting 1 | Indigenous or Pacific Islander | +0.4 (+0.1 to +0.6) |
| jw_no_tf | Setting 1 | Mixed | -0.2 (-0.5 to +0.1) |
| jw_no_tf | Setting 1 | Non-Hispanic Black | +0.1 (+0.0 to +0.1) |
| jw_no_tf | Setting 1 | Non-Hispanic White | +0.0 (+0.0 to +0.0) |
| jw_no_tf | Setting 1 | Other | -1.4 (-1.5 to -1.4) |
| jw_no_tf | Setting 1 | Unknown | -1.4 (-1.4 to -1.3) |
| levenshtein | Setting 1 | Asian | -0.8 (-0.8 to -0.7) |
| levenshtein | Setting 1 | Hispanic (White or Black) | -0.2 (-0.4 to -0.1) |
| levenshtein | Setting 1 | Indigenous or Pacific Islander | +0.5 (+0.2 to +0.7) |
| levenshtein | Setting 1 | Mixed | +0.2 (+0.1 to +0.3) |
| levenshtein | Setting 1 | Non-Hispanic Black | +0.0 (-0.0 to +0.1) |
| levenshtein | Setting 1 | Non-Hispanic White | +0.0 (+0.0 to +0.0) |
| levenshtein | Setting 1 | Other | -0.6 (-0.7 to -0.6) |
| levenshtein | Setting 1 | Unknown | -0.6 (-0.7 to -0.6) |
| levenshtein_no_tf | Setting 1 | Asian | -1.6 (-1.8 to -1.5) |
| levenshtein_no_tf | Setting 1 | Hispanic (White or Black) | -0.9 (-1.0 to -0.7) |
| levenshtein_no_tf | Setting 1 | Indigenous or Pacific Islander | +0.3 (+0.0 to +0.6) |
| levenshtein_no_tf | Setting 1 | Mixed | -0.2 (-0.5 to +0.1) |
| levenshtein_no_tf | Setting 1 | Non-Hispanic Black | +0.0 (-0.2 to +0.2) |
| levenshtein_no_tf | Setting 1 | Non-Hispanic White | +0.0 (+0.0 to +0.0) |
| levenshtein_no_tf | Setting 1 | Other | -1.4 (-1.5 to -1.3) |
| levenshtein_no_tf | Setting 1 | Unknown | -1.4 (-1.5 to -1.3) |
| combined | Setting 2 | Asian | -5.3 (-5.5 to -5.2) |
| combined | Setting 2 | Hispanic (White or Black) | +0.0 (-0.3 to +0.4) |
| combined | Setting 2 | Indigenous or Pacific Islander | +5.4 (+4.8 to +5.9) |
| combined | Setting 2 | Mixed | +5.0 (+4.7 to +5.3) |
| combined | Setting 2 | Non-Hispanic Black | -0.8 (-1.0 to -0.7) |
| combined | Setting 2 | Non-Hispanic White | +0.0 (+0.0 to +0.0) |
| combined | Setting 2 | Other | -3.5 (-3.7 to -3.3) |
| combined | Setting 2 | Unknown | -4.4 (-4.5 to -4.3) |
| jw | Setting 2 | Asian | -0.7 (-0.8 to -0.7) |
| jw | Setting 2 | Hispanic (White or Black) | -0.3 (-0.5 to -0.1) |
| jw | Setting 2 | Indigenous or Pacific Islander | +0.6 (+0.3 to +0.8) |
| jw | Setting 2 | Mixed | +0.1 (-0.2 to +0.3) |
| jw | Setting 2 | Non-Hispanic Black | +0.0 (-0.0 to +0.1) |
| jw | Setting 2 | Non-Hispanic White | +0.0 (+0.0 to +0.0) |
| jw | Setting 2 | Other | -0.6 (-0.7 to -0.5) |
| jw | Setting 2 | Unknown | -0.7 (-0.8 to -0.7) |
| jw_no_tf | Setting 2 | Asian | -1.7 (-1.8 to -1.6) |
| jw_no_tf | Setting 2 | Hispanic (White or Black) | -1.2 (-1.4 to -1.0) |
| jw_no_tf | Setting 2 | Indigenous or Pacific Islander | +0.6 (+0.2 to +1.1) |
| jw_no_tf | Setting 2 | Mixed | -0.3 (-0.5 to -0.1) |
| jw_no_tf | Setting 2 | Non-Hispanic Black | +0.1 (-0.0 to +0.1) |
| jw_no_tf | Setting 2 | Non-Hispanic White | +0.0 (+0.0 to +0.0) |
| jw_no_tf | Setting 2 | Other | -1.6 (-1.7 to -1.5) |
| jw_no_tf | Setting 2 | Unknown | -1.6 (-1.7 to -1.5) |
| levenshtein | Setting 2 | Asian | -0.9 (-0.9 to -0.8) |
| levenshtein | Setting 2 | Hispanic (White or Black) | -0.5 (-0.7 to -0.4) |
| levenshtein | Setting 2 | Indigenous or Pacific Islander | +0.6 (+0.4 to +0.7) |
| levenshtein | Setting 2 | Mixed | -0.1 (-0.3 to +0.1) |
| levenshtein | Setting 2 | Non-Hispanic Black | -0.0 (-0.0 to +0.0) |
| levenshtein | Setting 2 | Non-Hispanic White | +0.0 (+0.0 to +0.0) |
| levenshtein | Setting 2 | Other | -0.8 (-0.9 to -0.7) |
| levenshtein | Setting 2 | Unknown | -0.8 (-0.9 to -0.8) |
| levenshtein_no_tf | Setting 2 | Asian | -1.7 (-1.8 to -1.6) |
| levenshtein_no_tf | Setting 2 | Hispanic (White or Black) | -1.2 (-1.4 to -1.0) |
| levenshtein_no_tf | Setting 2 | Indigenous or Pacific Islander | +0.5 (-0.1 to +1.1) |
| levenshtein_no_tf | Setting 2 | Mixed | -0.4 (-0.7 to -0.2) |
| levenshtein_no_tf | Setting 2 | Non-Hispanic Black | -0.0 (-0.1 to +0.0) |
| levenshtein_no_tf | Setting 2 | Non-Hispanic White | +0.0 (+0.0 to +0.0) |
| levenshtein_no_tf | Setting 2 | Other | -1.5 (-1.6 to -1.3) |
| levenshtein_no_tf | Setting 2 | Unknown | -1.5 (-1.6 to -1.4) |
| combined | Setting 3 | Asian | +5.9 (+5.6 to +6.2) |
| combined | Setting 3 | Hispanic (White or Black) | +14.7 (+14.1 to +15.3) |
| combined | Setting 3 | Indigenous or Pacific Islander | +35.2 (+34.3 to +36.2) |
| combined | Setting 3 | Mixed | +19.5 (+18.9 to +20.2) |
| combined | Setting 3 | Non-Hispanic Black | +4.4 (+4.2 to +4.5) |
| combined | Setting 3 | Non-Hispanic White | +0.0 (+0.0 to +0.0) |
| combined | Setting 3 | Other | +14.2 (+13.9 to +14.5) |
| combined | Setting 3 | Unknown | +2.1 (+2.0 to +2.2) |
| jw | Setting 3 | Asian | -0.2 (-0.3 to -0.1) |
| jw | Setting 3 | Hispanic (White or Black) | +0.2 (+0.0 to +0.3) |
| jw | Setting 3 | Indigenous or Pacific Islander | +1.9 (+1.4 to +2.4) |
| jw | Setting 3 | Mixed | +0.5 (+0.4 to +0.6) |
| jw | Setting 3 | Non-Hispanic Black | +0.2 (+0.1 to +0.2) |
| jw | Setting 3 | Non-Hispanic White | +0.0 (+0.0 to +0.0) |
| jw | Setting 3 | Other | +0.0 (-0.1 to +0.1) |
| jw | Setting 3 | Unknown | -0.2 (-0.3 to -0.2) |
| jw_no_tf | Setting 3 | Asian | -1.5 (-1.6 to -1.4) |
| jw_no_tf | Setting 3 | Hispanic (White or Black) | -1.0 (-1.2 to -0.9) |
| jw_no_tf | Setting 3 | Indigenous or Pacific Islander | +3.7 (+3.2 to +4.3) |
| jw_no_tf | Setting 3 | Mixed | -0.6 (-1.0 to -0.2) |
| jw_no_tf | Setting 3 | Non-Hispanic Black | +0.1 (+0.1 to +0.2) |
| jw_no_tf | Setting 3 | Non-Hispanic White | +0.0 (+0.0 to +0.0) |
| jw_no_tf | Setting 3 | Other | -1.2 (-1.4 to -1.1) |
| jw_no_tf | Setting 3 | Unknown | -1.3 (-1.4 to -1.2) |
| levenshtein | Setting 3 | Asian | -0.2 (-0.3 to -0.2) |
| levenshtein | Setting 3 | Hispanic (White or Black) | +0.1 (-0.1 to +0.2) |
| levenshtein | Setting 3 | Indigenous or Pacific Islander | +1.9 (+1.5 to +2.4) |
| levenshtein | Setting 3 | Mixed | +0.3 (+0.2 to +0.5) |
| levenshtein | Setting 3 | Non-Hispanic Black | +0.2 (+0.1 to +0.2) |
| levenshtein | Setting 3 | Non-Hispanic White | +0.0 (+0.0 to +0.0) |
| levenshtein | Setting 3 | Other | -0.1 (-0.1 to -0.0) |
| levenshtein | Setting 3 | Unknown | -0.2 (-0.3 to -0.2) |
| levenshtein_no_tf | Setting 3 | Asian | -1.5 (-1.6 to -1.4) |
| levenshtein_no_tf | Setting 3 | Hispanic (White or Black) | -1.0 (-1.2 to -0.9) |
| levenshtein_no_tf | Setting 3 | Indigenous or Pacific Islander | +3.7 (+3.1 to +4.3) |
| levenshtein_no_tf | Setting 3 | Mixed | -0.7 (-1.2 to -0.2) |
| levenshtein_no_tf | Setting 3 | Non-Hispanic Black | +0.1 (-0.0 to +0.2) |
| levenshtein_no_tf | Setting 3 | Non-Hispanic White | +0.0 (+0.0 to +0.0) |
| levenshtein_no_tf | Setting 3 | Other | -1.2 (-1.4 to -1.1) |
| levenshtein_no_tf | Setting 3 | Unknown | -1.3 (-1.4 to -1.3) |

Supplementary Table 6. White-centric disparities in false match rate (FMR) by model, setting, and ethnic group (percentage points; mean and 95% CI across five replicates).